# Construction of potential functions associated with a given energy spectrum - An inverse problem. II

Abdulaziz D. Alhaidari[(a)†] and Houcine Aounallah[(b)]

[(a)] *Saudi Center for Theoretical Physics, P.O. Box 32741, Jeddah 21438, Saudi Arabia*

[(b)] *Department of Science and Technology, Larbi Tebessi University, 12000 Tebessa, Algeria*

**Abstract:** We continue our solution of the inverse problem started by the first author in [Int. J. Mod. Phys. A **35**, xxxx (2020)]. Additional potential functions for exactly solvable problems that correspond to the same energy spectrum formula but for different energy polynomials and bases are found. In this work, we obtain a class of potential functions associated with the Wilson polynomial and "Jacobi basis".

## 1. Introduction

Recently, we presented a solution of the inverse problem whereby potential functions associated with a given energy spectrum were constructed [1]. These potential functions were obtained numerically. In all cases, however, perfect fit to analytic functions were established. Some of these potentials correspond to a new class of exactly solvable problems that includes the one-dimensional logarithmic potential and the three-dimensional Coulomb plus linear potential. These do not belong to the conventional class of exactly solvable potentials. They are associated with the bound states energy spectrum formula $E_k = -\frac{1}{2}\lambda^2(k+\mu)^2$ where $\lambda$ is a scale parameter and $\mu$ is a dimensionless parameter. To pose the problem and obtain its solution, we used a formulation of quantum mechanics built not on potential functions but rather on orthogonal polynomials in the energy and physical parameters [2-5]. All physical properties of the system in this formulation are obtained from the properties of the polynomials. The total space-time wavefunction in this formulation is written as follows

$$\Psi(x,t) = \int e^{-iEt}\psi_E(x)dE + \sum_k e^{-iE_k t}\psi_k(x), \qquad (1)$$

where the continuous and discrete components of the wavefunction satisfy the time-independent wave equation $H\psi_E = E\psi_E$ and $H\psi_k = E_k\psi_k$, where $H$ is the Hamiltonian operator. If the system consists only of continuous (discrete) scattering (bound) states, then the sum (integral) part of the wavefunction is absent. The complete set of basis functions in configuration space, $\{\phi_n(x)\}$, is chosen properly such that we can write

---

[†] Corresponding Author. Email: haidari@sctp.org.sa

$$\psi_E(x) = \sqrt{\sigma(E)} \sum_n P_n(s) \phi_n(x), \tag{2a}$$

$$\psi_k(x) = \sqrt{\omega(s_k)} \sum_n P_n(s_k) \phi_n(x), \tag{2b}$$

where $\{P_n(s)\}$ are orthogonal polynomials whose argument $s$ is related to the energy by the measure relation $\sigma(E)dE = \rho(s)ds$ where $\rho(s)$ is the continuous part of the weight function of $\{P_n(s)\}$ and $\omega(s_k)$ is the discrete part. The polynomial sequence used in our previous work [1] and will also be used here has either a purely continuous spectrum or a mix of continuous and discrete spectrum where the polynomials satisfy the following normalized orthogonality relation

$$\int \rho(s) P_n(s) P_m(s) ds + \sum_k \omega(s_k) P_n(s_k) P_m(s_k) = \delta_{n,m}. \tag{3}$$

They also satisfy the following symmetric three-term recursion relation

$$s P_n(s) = a_n P_n(s) + b_{n-1} P_{n-1}(s) + b_n P_{n+1}(s), \tag{4}$$

where $n = 1,2,3,...$ The coefficients $\{a_n, b_n\}$ depend on $n$ and the physical parameters but are independent of $s$ and such that $b_n^2 > 0$ for all $n$. The asymptotics ($n \to \infty$) of $P_n(s)$ is a sinusoidal function in $n$ and it gives the spectrum of $P_n(s)$ as the discrete set $\{s_k\}$ at which the amplitude of the sinusoidal oscillation vanishes. The scattering phase shift is obtained from the argument of the sinusoidal function [2-5]. Substituting the wavefunction expansion (2) in the time-independent wave equation $H|\psi_E\rangle = E|\psi_E\rangle$ and projecting from left by $\langle\phi_n|$, we obtain the matrix wave equation $H|P\rangle = E|P\rangle$, where $H$ is the matrix representing the Hamiltonian in the basis $\{\phi_n\}$, which is assumed to be orthonormal (i.e., $\langle\phi_n|\phi_m\rangle = \delta_{n,m}$).

In our previous work, we have chosen $P_n(s)$ as the continuous dual Hah polynomial $S_n^\mu(z^2;a,b)$ whose spectrum formula is $z_k^2 = -(k+\mu)^2$ with $s = z^2$ and $k = 0,1,..,N$ where $N$ is the largest integer less than or equal to $-\mu$. In this work, however, we choose $P_n(s)$ as the Wilson polynomial $W_n(z^2;a,b,c,d)$ whose spectrum formula is the same, $z_k^2 = -(k+a)^2$, and $s = z^2$. The orthonormal version of this polynomial reads as follows (see Appendix C in [5])

$$W_n(z^2;a,b,c,d) = \sqrt{\left(\frac{2n+a+b+c+d-1}{n+a+b+c+d-1}\right)\frac{(a+b)_n(a+c)_n(a+d)_n(a+b+c+d)_n}{(b+c)_n(b+d)_n(c+d)_n n!}} \\ \times {}_4F_3\left({-n,n+a+b+c+d-1,a+iz,a-iz \atop a+v,a+c,a+d}\bigg|1\right) \tag{5}$$

where ${}_4F_3\left({a,b,c,d \atop e,f,g}\bigg|x\right) = \sum_{n=0}^{\infty} \frac{(a)_n(b)_n(c)_n(d)_n}{(e)_n(f)_n(g)_n}\frac{x^n}{n!}$ is the generalized hypergeometric function and $(a)_n$ is the Pochhammer symbol (a.k.a. shifted factorial): $a(a+1)(a+2)...(a+n-1) = \frac{\Gamma(n+a)}{\Gamma(a)}$. Our notation here for the Wilson polynomial is related to that in Appendix C of [5] by the map $W_n(z^2;a,b,c,d) \mapsto W_n^a(z^2;b,c,d)$. If the polynomial parameters are such that $\text{Re}(a,b,c,d) > 0$ with non-real parameters occurring in conjugate pairs then $W_n(z^2;a,b,c,d)$ is a polynomial of

degree $n$ in $z^2$ with a purely continuous spectrum on the positive real line, $z \geq 0$. In this case, the orthogonality of this polynomial sequence reads as follows (see Appendix C in [5])

$$\int_0^\infty \rho(z) W_n(z^2;a,b,c,d) W_m(z^2;a,b,c,d) \, dz = \delta_{nm}, \quad \text{where} \tag{6a}$$

$$\rho(z) = \frac{1}{2\pi} \frac{\Gamma(a+b+c+d)\left|\Gamma(a+iz)\Gamma(b+iz)\Gamma(c+iz)\Gamma(d+iz)/\Gamma(2iz)\right|^2}{\Gamma(a+b)\Gamma(c+d)\Gamma(a+c)\Gamma(a+d)\Gamma(b+c)\Gamma(b+d)}. \tag{6b}$$

On the other hand, if the parameters are such that $a < 0$ and $\text{Re}(a+b, a+c, a+d) > 0$ with non-real parameters occurring in conjugate pairs, then the polynomial will have a continuous spectrum on $z \geq 0$ as well as a finite size discrete spectrum on $z < 0$. In this case, the polynomial sequence satisfies a generalized orthogonality relation similar to (3) and given by Eq. (C3) in [5] where the discrete weight function reads as follows

$$\omega(z_k) = -2\frac{\Gamma(a+b+c+d)\Gamma(b-a)\Gamma(c-a)\Gamma(d-a)}{\Gamma(-2a+1)\Gamma(c+d)\Gamma(b+c)\Gamma(b+d)}\left[(k+a)\frac{(2a)_k(a+b)_k(a+c)_k(a+d)_k}{(a-b+1)_k(a-c+1)_k(a-d+1)_k k!}\right]. \tag{7}$$

The symmetric three-term recursion relation satisfied by $W_n(z^2;a,b,c,d)$ is given by Eq. (C4) in Appendix C of [5]. Comparing that to Eq. (4) above, which could be written in matrix form as $s|P\rangle = \Sigma|P\rangle$, gives the following elements of the tridiagonal symmetric matrix $\Sigma$

$$\Sigma_{n,m} = \left[\frac{(n+a+b)(n+a+c)(n+a+d)(n+a+b+c+d-1)}{(2n+a+b+c+d)(2n+a+b+c+d-1)} + \frac{n(n+b+c-1)(n+b+d-1)(n+c+d-1)}{(2n+a+b+c+d-1)(2n+a+b+c+d-2)} - a^2\right]\delta_{n,m}$$
$$- \frac{1}{2n+a+b+c+d-2}\sqrt{\frac{n(n+a+b-1)(n+c+d-1)(n+a+c-1)(n+a+d-1)(n+b+c-1)(n+b+d-1)(n+a+b+c+d-2)}{(2n+a+b+c+d-3)(2n+a+b+c+d-1)}}\delta_{n,m+1} \tag{8}$$
$$- \frac{1}{2n+a+b+c+d}\sqrt{\frac{(n+1)(n+a+b)(n+c+d)(n+a+c)(n+a+d)(n+b+c)(n+b+d)(n+a+b+c+d-1)}{(2n+a+b+c+d-1)(2n+a+b+c+d+1)}}\delta_{n,m-1}$$

Comparing the matrix wave equation $E|P\rangle = H|P\rangle$ to $z^2|P\rangle = \Sigma|P\rangle$ and the spectrum formula of the polynomial $z_k^2 = -(k+a)^2$ to the energy spectrum $E_k = -\frac{1}{2}\lambda^2(k+a)^2$, we conclude that the energy is related to $z$ by $E = \frac{1}{2}\lambda^2 z^2$ and thus the matrix representation of the Hamiltonian operator is $H = \frac{1}{2}\lambda^2\Sigma$.

Now, for a full description of the system whose wavefunction is given by Eq. (1) and Eq. (2), we need to specify the basis set $\{\phi_n(x)\}$. In our previous work, we chose the "Laguerre basis" whose elements are written in terms of the Laguerre polynomial [1]. In this work, however, we choose the "Jacobi basis" whose elements are

$$\phi_n(x) = A_n(1-y)^\alpha(1+y)^\beta P_n^{(\mu,\nu)}(y), \tag{9}$$

where $P_n^{(\mu,\nu)}(y)$ is the Jacobi polynomial with real parameters such that $(\mu,\nu) > -1$ and $A_n$ is a proper normalization constant. The dimensionless argument of the polynomial is a coordinate transformation such that $-1 \leq y(x) \leq +1$ and $dy/dx = \lambda\gamma(1-y)^\eta(1+y)^\tau$. Table 1 gives a list of the parameters $\{\eta,\tau,\gamma\}$ for the basis sets that correspond to the sample problems treated in section 2 below and with $2\alpha = \mu + \eta$ and $2\beta = \nu + \tau$. The rest of the parameters, $\{\lambda,\mu,\nu\}$, are related to the physical parameters of the corresponding problem. To verify orthonormality of the sets, one uses the orthogonality of the Jacobi polynomials,

–3–

$$\int_{-1}^{+1} (1-y)^\mu (1+y)^\nu P_n^{(\mu,\nu)}(y) P_m^{(\mu,\nu)}(y) dy = \frac{2^{\mu+\nu+1}}{2n+\mu+\nu+1} \frac{\Gamma(n+\mu+1)\Gamma(n+\nu+1)}{\Gamma(n+1)\Gamma(n+\mu+\nu+1)} \delta_{n,m}, \quad (10)$$

in addition to the integral measure $\int_{x_-}^{x_+} dx = \int_{-1}^{+1} \frac{dy}{y'}$, where $x_\pm$ are the boundaries of configuration space and $y' = dy/dx$.

In Appendix A, we show how to derive the symmetric matrix $T$ representing the kinetic energy operator in the Jacobi bases (9). Consequently, the potential matrix is obtained as $V = H - T$ with $H$ being given above as $\frac{1}{2}\lambda^2 \Sigma$. In section 2, we use a procedure to construct the potential function using its matrix elements and the corresponding basis set. This procedure was outlined in Appendix A of our previous work [1]. We end in section 3 with a conclusion and discussion of our findings.

## 2. Illustrative examples

In the following subsections, we present five examples of systems whose wavefunctions are given by Eq. (1) and Eq. (2) and where the energy polynomial is $P_n(z) = W_n(z^2; a, b, c, d)$. The elements of the basis $\{\phi_n(x)\}$ are given by Eq. (9) with the parameters shown in the entries of Table 1. In all the examples, we choose $\lambda = 1.0$ and $a = -4.5$ but vary $b$ and $c$ while keeping $d = c$, $\mu = 2c - 1$ and $\nu = a + b - 1$. Similar to our previous work [1], the solution obtained is not unique. The non-uniqueness shows up here as arbitrary shifting and scaling of the resulting potential function. That is, the potential is unique modulo two "equivalence parameters" $q_0$ and $q_1$ where the derived potential $\hat{V}(x)$ is related to the true potential as $V(x) = q_0 \left[ \hat{V}(x) - q_1 \right]$. These two parameters depend on the physical parameters of the problem but $q_0$ must be positive. In most cases, $q_1$ is fixed by the requirement that $V(x)$ vanish at infinity. That is, $q_1 = \hat{V}(\infty)$. On the other hand, only a lower bound on $q_0$ could be obtained for the cases with bound states. This is the condition that $q_0 \left[ \hat{V}(\infty) - \hat{V}(x_0) \right] > \frac{1}{2}(\lambda a)^2$, where $x_0$ is the location of the absolute (global) minimum of the potential.

The matrix representation of the kinetic energy is obtained by calculating the action of the kinetic energy differential operator on the elements of the basis as $T | \phi_m \rangle$. Subsequently, the matrix elements are obtained by projecting from left with $\langle \phi_n |$ then evaluating the integral as $T_{n,m} = \langle \phi_n | T | \phi_m \rangle$. This calculation is carried out in Appendix A. In the process of calculating the matrix elements $\langle \phi_n | T | \phi_m \rangle$, we encounter integrals of the form

$$\frac{A_n A_m}{\gamma} \int_{-1}^{+1} (1-y)^{\mu+\alpha} (1+y)^{\nu+\beta} P_n^{(\mu,\nu)}(y) P_m^{(\mu,\nu)}(y) dy := F_{n,m}^{(\mu,\nu)}(\alpha, \beta), \quad (11)$$

where $A_n = \sqrt{\gamma} \sqrt{\frac{2n+\mu+\nu+1}{2^{\mu+\nu+1}} \frac{\Gamma(n+1)\Gamma(n+\mu+\nu+1)}{\Gamma(n+\mu+1)\Gamma(n+\nu+1)}}$ and the real parameters are such that $\mu > -1$, $\nu > -1$, $\mu + \alpha > -1$ and $\nu + \beta > -1$. This integral is evaluated in Appendix B.

The matrix representation of the kinetic energy given by Eq. (A8) in Appendix A suggests that we can absorb all terms with constant (*n*-independent) coefficients multiplying $F_{m,n}^{(\mu,\nu)}(\alpha,\beta)$ into the potential function by writing it as

$$V(x) = \frac{(1-y)^{2\eta}(1+y)^{2\tau}}{1-y^2}\left[V_0 + \frac{V_+}{1+y} + \frac{V_-}{1-y}\right] + \tilde{V}(x), \tag{12}$$

and choosing the basis parameters in terms of these potential parameters as follows

$$-\frac{4V_0}{(\gamma\lambda)^2} = (\mu+\nu+1)(\eta+\tau-1) + \frac{3}{4}(\eta+\tau-1)^2, \tag{13a}$$

$$-\frac{4V_+}{(\gamma\lambda)^2} = \frac{1}{2}\left[(\tau-1)^2 - (\nu+2\tau-1)^2\right], \tag{13b}$$

$$-\frac{4V_-}{(\gamma\lambda)^2} = \frac{1}{2}\left[(\eta-1)^2 - (\mu+2\eta-1)^2\right]. \tag{13c}$$

Then, the Hamiltonian, $H = T + V$, becomes $H = \tilde{T} + \tilde{V}$ where the kinetic energy matrix $\tilde{T}$ corresponding to the potential component $\tilde{V}(x)$ becomes

$$-\frac{4}{(\gamma\lambda)^2}\langle\phi_m|\tilde{T}|\phi_n\rangle = -\left(n + \frac{\mu+\nu+1}{2}\right)^2 F_{m,n}^{(\mu,\nu)}(2\eta-1, 2\tau-1)$$
$$+ 2(\mu-\nu)G_n\left[(2\tau-1)F_{m,n}^{(\mu,\nu)}(2\eta-1, 2\tau-2) - (2\eta-1)F_{m,n}^{(\mu,\nu)}(2\eta-2, 2\tau-1)\right]$$
$$+ (n+\mu+\nu+1)D_{n-1}\left[(2\tau-1)F_{m,n-1}^{(\mu,\nu)}(2\eta-1, 2\tau-2) - (2\eta-1)F_{m,n-1}^{(\mu,\nu)}(2\eta-2, 2\tau-1)\right] \tag{14}$$
$$- nD_n\left[(2\tau-1)F_{m,n+1}^{(\mu,\nu)}(2\eta-1, 2\tau-2) - (2\eta-1)F_{m,n+1}^{(\mu,\nu)}(2\eta-2, 2\tau-1)\right]$$
$$+ n \leftrightarrow m$$

An alternative and simpler expression could be obtained by using the differential property of the Jacobi polynomial (A2b) where we can show that the right-hand side of Eq. (14), without the first term, is equal to $(2\tau-1)\tilde{F}_{m,n}^{(\mu,\nu)}(2\eta, 2\tau-1) - (2\eta-1)\tilde{F}_{m,n}^{(\mu,\nu)}(2\eta-1, 2\tau)$ where

$$\tilde{F}_{n,m}^{(\mu,\nu)}(\alpha,\beta) := \frac{A_n A_m}{\gamma}\int_{-1}^{+1}(1-y)^{\mu+\alpha}(1+y)^{\nu+\beta}P_n^{(\mu,\nu)}(y)\left[\frac{d}{dy}P_m^{(\mu,\nu)}(y)\right]dy, \tag{15}$$

This integral is evaluated in Appendix B and is given by Eq. (B10). Therefore, we can rewrite Eq. (14) in the following simpler but equivalent form

$$\frac{4}{(\gamma\lambda)^2}\langle\phi_m|\tilde{T}|\phi_n\rangle = \left(n + \frac{\mu+\nu+1}{2}\right)^2 F_{m,n}^{(\mu,\nu)}(2\eta-1, 2\tau-1)$$
$$-(2\tau-1)\tilde{F}_{m,n}^{(\mu,\nu)}(2\eta, 2\tau-1) + (2\eta-1)\tilde{F}_{m,n}^{(\mu,\nu)}(2\eta-1, 2\tau) + n \leftrightarrow m \tag{16}$$

For some cases, this alternative expression could be numerically more stable and/or rapidly convergent than (14). In the special case where $\eta + \tau = 1$, it simplifies even further to read

$$\frac{4}{(\gamma\lambda)^2}\langle\phi_m|\tilde{T}|\phi_n\rangle = \left(n+\frac{\mu+\nu+1}{2}\right)^2 F_{m,n}^{(\mu,\nu)}(2\eta-1,2\tau-1)$$
$$+2(\eta-\tau)\tilde{F}_{m,n}^{(\mu,\nu)}(2\eta-1,2\tau-1)+n\leftrightarrow m \tag{17}$$

Now, with knowledge of the matrix $\tilde{T}$ and the basis (9), the construction technique of the potential function described in Appendix A of [1] shifts from $V(x)$ to the potential component $\tilde{V}(x)$ for a selected set of physical parameters. Finally, we add $\tilde{V}(x)$ to the rest of the potential terms as shown in Eq. (12) to obtain $V(x)$.

**2.1 The case** $y(x)=2\tanh^2(\lambda x)-1$ and $x\geq 0$:

This case corresponds to the 1D problem on the positive real line whose basis parameters are shown in the first row of Table 1. Using these parameters in Eq. (12), we obtain the following potential function

$$V(x)=V_- + \frac{2V_0}{\cosh^2(\lambda x)} + \frac{V_+}{\sinh^2(\lambda x)} + \tilde{V}(x). \tag{18}$$

To derive the potential component $\tilde{V}(x)$, we use the same parameters and substitute in Eq. (14) to obtain the following elements of the corresponding kinetic energy matrix

$$\frac{1}{\lambda^2}\langle\phi_m|\tilde{T}|\phi_n\rangle = \left[\left(n+\frac{\mu+\nu+1}{2}\right)^2(1-C_n)+2(\mu-\nu)G_n\right]\delta_{m,n}$$
$$-\left[\left(n+\frac{\mu+\nu}{2}\right)^2+\frac{1}{16}\right]D_{n-1}\delta_{m,n-1}-\left[\left(n+1+\frac{\mu+\nu}{2}\right)^2+\frac{1}{16}\right]D_n\delta_{m,n+1} \tag{19}$$
$$+\frac{1}{2}(\mu+\nu+2)\left(D_{n-1}\delta_{m,n-1}+D_n\delta_{m,n+1}\right)$$

where the coefficients $C_n$ and $D_n$ are defined below Eq. (B2) in Appendix B whereas the coefficient $G_n$ is defined below Eq. (A8) in Appendix A. The matrix $\tilde{T}$ above and the Hamiltonian matrix $H=\frac{1}{2}\lambda^2\Sigma$ give the potential matrix as $\tilde{V}=H-\tilde{T}$. Using this potential matrix and the corresponding basis, the method outlined in Appendix A of [1] gives the potential function $\tilde{V}(x)$ shown as solid line in Fig. 1a for a given set of physical parameters $b$ and $c$. By empirical trials, we were able to obtain a perfect fit[‡] to the potential function $\tilde{V}(x)=\tilde{V}_0+\frac{\tilde{V}_1}{\cosh^2(\lambda x)}$ for a given parameters $\tilde{V}_0$ and $\tilde{V}_1$ as shown by the dotted curve in Fig. 1a. Consequently, our result is identical to the exactly solvable Pöschl-Teller potential (see, for example, the treatment in section 5.1 in Ref. [5]). In figure 1b, we plot the total potential (18) for a give equivalence parameters $\{q_0,q_1\}$ chosen as explained above. We also superimpose the energy spectrum levels on the same plot.

**2.2 The case** $y(x)=\tanh(\lambda x)$ and $-\infty<x<+\infty$:

---

[‡] In this work and in [1], the term "perfect fit" means that the potential function and the fitting function match to within the computing machine accuracy (e.g., for single precision calculation, local differences between the two functions are in the order of $10^{-16}$ to $10^{-14}$).

This case corresponds to a 1D problem on the whole real line with basis parameters as shown in the second row of Table 1. Using these parameters in Eq. (12), we obtain the following potential function

$$V(x) = W_+ - W_- \tanh(\lambda x) + \frac{V_0}{\cosh^2(\lambda x)} + \tilde{V}(x), \qquad (20)$$

where $W_\pm = V_+ \pm V_-$. To derive the potential component $\tilde{V}(x)$, we use the same parameters and substitute in Eq. (14) to obtain the following elements of the corresponding kinetic energy matrix

$$-\frac{4}{\lambda^2}\langle \phi_m | \tilde{T} | \phi_n \rangle = -\left(n + \frac{\mu+\nu+1}{2}\right)^2 \left(\delta_{m,n} - K^2_{m,n}\right)$$
$$-4(\mu-\nu)G_n K_{m,n} - 2(n+\mu+\nu+1)D_{n-1}K_{m,n-1} + 2nD_n K_{m,n+1} \qquad (21)$$
$$+ n \leftrightarrow m$$

where the tridiagonal symmetric matrix $K$ is defined by Eq. (B2) in Appendix B. The coefficient $G_n$ is defined in Appendix A whereas the coefficients $D_n$ is defined in Appendix B. The kinetic energy matrix $\tilde{T}$ above and the Hamiltonian matrix $H = \frac{1}{2}\lambda^2 \Sigma$ give the potential matrix as $\tilde{V} = H - \tilde{T}$. Using this potential matrix and the corresponding basis, the method outlined in Appendix A of [1] gives the potential function $\tilde{V}(x)$ shown as solid line in Fig. 2a for a given set of physical parameters $b$ and $c$. By empirical trials, we were able to obtain a perfect fit to the potential function $\tilde{V}(x) = \tilde{V}_0 + \tilde{V}_1 \tanh(\lambda x) + \frac{\tilde{V}_2}{\cosh^2(\lambda x)}$ for a given set of parameters $\{\tilde{V}_i\}$ as shown by the dotted curve in Fig. 2a. Therefore, the total potential (20) becomes the hyperbolic Rosen-Morse potential [6], which is plotted for a proper choice of the equivalence parameters $\{q_0, q_1\}$ in Fig. 2b where we also superimpose the energy spectrum levels.

**2.3 The case** $y(x) = 1 - 2e^{-\lambda x}$ **and** $x \geq 0$:

This case corresponds to the 1D problem on the positive real line whose basis parameters are shown in the third row of Table 1. Using these parameters in Eq. (12), we obtain the following potential function

$$V(x) = \frac{V_0}{e^{\lambda x} - 1} + \frac{1/2}{1 - e^{-\lambda x}}\left(V_- + \frac{V_+}{e^{\lambda x} - 1}\right) + \tilde{V}(x), \qquad (22)$$

To derive the potential component $\tilde{V}(x)$, we use the same parameters and substitute in Eq. (14) to obtain the following elements of the corresponding kinetic energy matrix

$$-\frac{4}{\lambda^2}\langle \phi_m | \tilde{T} | \phi_n \rangle = -\left(n + \frac{\mu+\nu+1}{2}\right)^2 F^{(\mu,\nu)}_{m,n}(1,-1) - 4(\mu-\nu)G_n F^{(\mu,\nu)}_{m,n}(0,-2)$$
$$-2(n+\mu+\nu+1)D_{n-1}F^{(\mu,\nu)}_{m,n-1}(0,-2) + 2nD_n F^{(\mu,\nu)}_{m,n+1}(0,-2) \qquad (23)$$
$$+ n \leftrightarrow m$$

where we have used the simple identity $\frac{1}{1+y} + \frac{1-y}{(1+y)^2} = \frac{2}{(1+y)^2}$ to write $F_{n,m}^{(\mu,\nu)}(0,-1) + F_{n,m}^{(\mu,\nu)}(1,-2) = 2F_{n,m}^{(\mu,\nu)}(0,-2)$. Since in this case $\eta + \tau = 1$, we can use the alternative and simpler expression (17) and write

$$\frac{4}{\lambda^2}\langle\phi_m|\tilde{T}|\phi_n\rangle = \left(n + \frac{\mu+\nu+1}{2}\right)^2 F_{m,n}^{(\mu,\nu)}(1,-1) + 2\tilde{F}_{m,n}^{(\mu,\nu)}(1,-1) + n \leftrightarrow m. \tag{24}$$

This matrix and the Hamiltonian matrix $H = \frac{1}{2}\lambda^2 \Sigma$ give the potential matrix as $\tilde{V} = H - \tilde{T}$. Using this potential matrix and the corresponding basis, the method described in Appendix A of [1] gives the potential function $\tilde{V}(x)$ shown in Fig. 3a for a given set of physical parameters $b$ and $c$. Unfortunately, in this case we were unable to produce accurate enough functional fit to $\tilde{V}(x)$. Nonetheless, the total potential function (22) is plotted in Fig. 3b for a proper choice of the equivalence parameters $\{q_0, q_1\}$ where we also superimpose the energy spectrum levels.

**2.4 The case** $y(r) = \frac{(\lambda r)^2 - 1}{(\lambda r)^2 + 1}$ **and** $r \geq 0$:

This case corresponds to the 3D problem with spherical symmetry whose basis parameters are shown in the fourth row of Table 1. Using these parameters in Eq. (A8) and Eq. (A9), we obtain the following elements of the matrix representation of the kinetic energy operators, $-\frac{1}{2}\frac{d^2}{dr^2}$ and $\frac{\ell(\ell+1)}{2r^2}$,

$$-\frac{4}{\lambda^2}\langle\phi_m|-\frac{1}{2}\frac{d^2}{dr^2}|\phi_n\rangle = -\left[\left(n + \frac{\mu+\nu+1}{2}\right)^2 + \frac{3}{4} + (\mu+\nu+1)\right]\left[(1-K)^2\right]_{m,n}$$
$$+\frac{1}{2}\left[(\mu+2)^2 - \frac{1}{4} - 8(\mu-\nu)G_n\right](\delta_{m,n} - K_{m,n}) + \frac{1}{2}\left(\nu^2 - \frac{1}{4}\right)F_{m,n}^{(\mu,\nu)}(2,-1) \tag{25a}$$
$$-2(n+\mu+\nu+1)D_{n-1}(\delta_{m,n-1} - K_{m,n-1}) + 2nD_n(\delta_{m,n+1} - K_{m,n+1})$$
$$+ n \leftrightarrow m$$

$$-\frac{4}{\lambda^2}\langle\phi_m|\frac{\ell(\ell+1)}{2r^2}|\phi_n\rangle = -2\ell(\ell+1)F_{m,n}^{(\mu,\nu)}(1,-1). \tag{25b}$$

Using the identities $\frac{1-y}{1+y} = \frac{2}{1+y} - 1$ and $\frac{(1-y)^2}{1+y} = \frac{4}{1+y} - (1-y) - 2$, one can show that the integral $F_{n,m}^{(\mu,\nu)}(\alpha,\beta)$ satisfies the following relation

$$F_{n,m}^{(\mu,\nu)}(1,-1) = 2F_{n,m}^{(\mu,\nu)}(0,-1) - \delta_{n,m}, \tag{26a}$$

$$F_{n,m}^{(\mu,\nu)}(2,-1) = 4F_{n,m}^{(\mu,\nu)}(0,-1) - F_{n,m}^{(\mu,\nu)}(1,0) - 2\delta_{n,m}. \tag{26b}$$

Therefore, choosing the basis parameter $\nu$ as $\nu = \ell + \frac{1}{2}$ simplifies the kinetic energy matrix, which is the sum of (25a) and (25b), to read as follows

$$-\frac{4}{\lambda^2}\langle\phi_m|T|\phi_n\rangle = -\left[\left(n+\frac{\mu+\nu+1}{2}\right)^2 + \frac{3}{4} + (\mu+\nu+1)\right]\left[(1-K)^2\right]_{m,n}$$
$$+\frac{1}{2}\left[(\mu+2)^2 - \left(\ell+\tfrac{1}{2}\right)^2 - 8(\mu-\nu)G_n\right]\left(\delta_{m,n} - K_{m,n}\right) \tag{27}$$
$$-2(n+\mu+\nu+1)D_{n-1}\left(\delta_{m,n-1} - K_{m,n-1}\right) + 2nD_n\left(\delta_{m,n+1} - K_{m,n+1}\right)$$
$$+ n \leftrightarrow m$$

This kinetic energy matrix and the Hamiltonian matrix $H = \frac{1}{2}\lambda^2 \Sigma$ give the potential matrix as $V = H - T$. The method described in Appendix A of [1] uses this potential matrix and the corresponding basis to construct the potential function (including the orbital term $\frac{\ell(\ell+1)}{2r^2}$) shown as solid line in Fig. 4 for a given set of physical parameters $\{b, c, \ell\}$ and equivalence parameters $\{q_0, q_1\}$ where we also superimpose the energy spectrum levels. By empirical trials, we were able to obtain a perfect fit (shown as the dotted line in the figure) to the radial potential function $V(r) = V_0 + \frac{V_1}{1+(\lambda r)^2} + \frac{V_2}{\left[1+(\lambda r)^2\right]^2}$ for a given set of parameters $V_0$, $V_1$ and $V_2$. This new potential function does not belong to the known class of exactly solvable potentials.

**2.5 The case** $y(x) = \sin(\lambda x)$ **and** $-\frac{1}{2}\pi \geq \lambda x \geq +\frac{1}{2}\pi$:

This case corresponds to the 1D potential box problem whose basis parameters are shown in the fifth row of Table 1. Using those parameters in Eq. (12), we obtain the following potential function

$$V(x) = V_0 + \frac{W_+ - W_- \sin(\lambda x)}{\cos^2(\lambda x)} + \tilde{V}(x), \tag{28}$$

where $W_\pm = V_+ \pm V_-$. To derive the potential component $\tilde{V}(x)$, we use the same parameters of the problem from the Table into Eq. (14) to obtain the following elements of the corresponding kinetic energy matrix

$$\langle\phi_m|\tilde{T}|\phi_n\rangle = \frac{\lambda^2}{2}\left(n + \frac{\mu+\nu+1}{2}\right)^2 \delta_{m,n}. \tag{29}$$

This matrix and $H = \frac{1}{2}\lambda^2 \Sigma$ give the potential matrix as $\tilde{V} = H - \tilde{T}$. Employing the method outlined in Appendix A of [1], we produce the potential function $\tilde{V}(x)$ shown as solid line in Fig. 5a for a given set of physical parameters $b$ and $c$. By empirical trials, we were able to obtain a perfect fit to the potential function $\tilde{V}(x) = \tilde{V}_0\left[1 - \sin(\lambda x)\right]$ for a given parameter $\tilde{V}_0$. This is shown as the dotted line in Fig. 5a. Therefore, the full potential function for this problem is given by Eq. (28) and reads as follows

$$V(x) = \begin{cases} V_0 + \dfrac{W_+ - W_- \sin(\lambda x)}{\cos^2(\lambda x)} + \tilde{V}_0\left[1 - \sin(\lambda x)\right] & ,|x| \leq \pi/2\lambda \\ \infty & ,\text{otherwise} \end{cases} \tag{30}$$

This potential function is shown in Fig. 5b with the equivalence parameters $q_0 = 1$ and $q_1 = 0$. It is identical to the exactly solvable potential treated in section III.A.1 of [7]. Without the $\tilde{V}(x)$ term, it becomes the usual trigonometric Scarf potential (see, for example, section 5.2 in Ref. [5]). It should be noted that (30) is a totally confining potential that corresponds to a system with purely discrete spectrum. Thus, its total set of bound states energies does not follow from the spectrum formula of the Wilson polynomial since this polynomial does not have a purely discrete spectrum. On the other hand, the total set of energy spectrum for this discrete system is obtained from another discrete orthogonal polynomial introduced in [8,9] and designated as $H_n^{(\mu,\nu)}(z;\alpha,\theta)$. Nonetheless, the system is completely defined and faithfully represented by its wavefunction, which is the discrete part of (1) with Fourier components as given by (2b).

## 4. Conclusion

The formulation of quantum mechanics, which was recently introduced in [2-5], has no reference at all to a potential function. The wavefunction is written as a bounded series of complete square integrable functions in configuration space. The expansion coefficients of the series are orthogonal polynomials in the energy and physical parameters. All properties of the physical system are obtained from those of the polynomials (e.g., their zeros, weight function, generating function, recursion relation, asymptotics, etc.). One advantage of the formulation is that the class of integrable systems (exactly solvable systems) is larger than the conventional class since one could include integrable systems whose potential functions may not be realized analytically.

To establish a correspondence between this formulation and the conventional one, we developed a technique to construct the potential function in this new formulation [1,10]. The technique consists of three steps. The first is to create the Hamiltonian matrix $H$ using the energy spectrum formula and the recursion relation of the orthogonal polynomials. The second step is to obtain the kinetic energy matrix $T$ using the give basis. The third and final step is to use the potential matrix $H - T$ and the basis to construct the potential function for a given set of physical parameters using one of four methods outlined in section 3 of [10]. As such, the work is viewed as a solution to the inverse problem. Note that in the conventional formulation of quantum mechanics, the inverse problem is posed as follows: structural and/or scattering information is provided and one is then asked to derive the associated potential function. However, in our alternative formulation and as explained in the introduction, this information is contained in the set of orthogonal polynomials $\{P_n(s)\}$. Therefore, the inverse problem is posed in the new formulation as follows: the set $\{P_n(s)\}$ is given and one is asked to derive the associated potential function. Moreover, we found out (as in the conventional formulation) that the solution obtained is not unique.

In this paper, we continued our work that was initiated in [1] and studied several systems but for different choices of orthogonal polynomial and basis. We specialized to the Wilson polynomial (5) and Jacobi basis (9). An interesting finding of our work here is a novel integrable system in 3D with spherical symmetry and with the following exactly solvable potential

$$V(r) = \frac{V_1}{1+(\lambda r)^2} + \frac{V_2}{\left[1+(\lambda r)^2\right]^2} . \tag{31}$$

This is a short-range regular potential that could be used as a model to describe the binding of various molecular systems. It remains a non-trivial exercise to relate the Wilson polynomial parameters $\{a,b,c\}$ to the potential parameters $\{V_1, V_2\}$ and $\ell$.

## Appendix A: The kinetic energy matrix in the Jacobi basis (9)

In all of the orthonormal Jacobi basis sets whose parameters are given in Table 1, the coordinate transformation $y(x)$ satisfies the following

$$\frac{1}{\lambda}\frac{dy}{dx} = \gamma(1-y)^{\eta}(1+y)^{\tau}, \tag{A1}$$

where the dimensionless parameters $\{\gamma, \eta, \tau\}$ are given in Table 1. The kinetic energy operator $T$ is a well-known differential operator in configuration space that depends only on the number of dimensions. For example, in one dimension with coordinate $x$ and adopting the atomic units $\hbar = m = 1$, $T = -\frac{1}{2}\frac{d^2}{dx^2}$. In three dimensions with spherical symmetry and radial coordinate $r$, $T = -\frac{1}{2}\frac{d^2}{dr^2} + \frac{\ell(\ell+1)}{2r^2}$ where $\ell$ is the angular momentum quantum number. Therefore, the action of $T$ on the given basis elements $\{\phi_n\}$ could be derived and so too its matrix elements. In this Appendix, we start by calculating the action of the differential operator part of $T$, $-\frac{1}{2}\frac{d^2}{dx^2}$, on the basis elements (9). For this calculation, one needs the following differential equation, differential property and recursion relation of the Jacobi polynomials

$$\left\{(1-y^2)\frac{d^2}{dy^2} - [(\mu+\nu+2)y + \mu - \nu]\frac{d}{dy} + n(n+\mu+\nu+1)\right\} P_n^{(\mu,\nu)}(y) = 0, \tag{A2a}$$

$$(1-y^2)\frac{d}{dy} P_n^{(\mu,\nu)}(y) = 2(n+\mu+\nu+1)\left[\frac{(\mu-\nu)n}{(2n+\mu+\nu)(2n+\mu+\nu+2)} P_n^{(\mu,\nu)}(y)\right.$$
$$\left. + \frac{(n+\mu)(n+\nu)}{(2n+\mu+\nu)(2n+\mu+\nu+1)} P_{n-1}^{(\mu,\nu)}(y) - \frac{n(n+1)}{(2n+\mu+\nu+1)(2n+\mu+\nu+2)} P_{n+1}^{(\mu,\nu)}(y)\right] \tag{A2b}$$

$$y P_n^{(\mu,\nu)}(y) = \frac{\nu^2 - \mu^2}{(2n+\mu+\nu)(2n+\mu+\nu+2)} P_n^{(\mu,\nu)}(y)$$
$$+ \frac{2(n+\mu)(n+\nu)}{(2n+\mu+\nu)(2n+\mu+\nu+1)} P_{n-1}^{(\mu,\nu)}(y) + \frac{2(n+1)(n+\mu+\nu+1)}{(2n+\mu+\nu+1)(2n+\mu+\nu+2)} P_{n+1}^{(\mu,\nu)}(y) \tag{A2c}$$

Using (A1) and the differentiation chain rule, we can write

$$\frac{1}{\lambda^2}\frac{d^2}{dx^2}|\phi_n(x)\rangle = \gamma^2(1-y)^{2\eta}(1+y)^{2\tau}\left[\frac{d^2}{dy^2} + \left(\frac{\tau}{1+y} - \frac{\eta}{1-y}\right)\frac{d}{dy}\right]|\phi_n(x)\rangle = \gamma^2 A_n (1-y)^{2\eta+\alpha}$$
$$\times (1+y)^{2\tau+\beta}\left[\left(\frac{d}{dy} + \frac{\beta}{1+y} - \frac{\alpha}{1-y}\right)^2 + \left(\frac{\tau}{1+y} - \frac{\eta}{1-y}\right)\left(\frac{d}{dy} + \frac{\beta}{1+y} - \frac{\alpha}{1-y}\right)\right]|P_n^{(\mu,\nu)}(y)\rangle \tag{A3}$$

Collecting terms leads to the following

$$\frac{1}{(\gamma\lambda)^2}\frac{d^2}{dx^2}|\phi_n(x)\rangle = A_n(1-y)^{2\eta+\alpha}(1+y)^{2\tau+\beta}\left[\frac{d^2}{dy^2}+\left(\frac{2\beta+\tau}{1+y}-\frac{2\alpha+\eta}{1-y}\right)\frac{d}{dy}\right.$$
$$\left.+\frac{\beta(\beta+\tau-1)}{(1+y)^2}+\frac{\alpha(\alpha+\eta-1)}{(1-y)^2}-\frac{2\alpha\beta+\alpha\tau+\beta\eta}{1-y^2}\right]|P_n^{(\mu,\nu)}(y)\rangle \quad (A4)$$

Using the differential equation of the Jacobi polynomials (A2a) and the identity $\frac{1\pm y}{1\mp y}=\frac{2}{1\mp y}-1$, we obtain

$$\frac{1}{(\gamma\lambda)^2}\frac{d^2}{dx^2}|\phi_n(x)\rangle = A_n(1-y)^{2\eta+\alpha-1}(1+y)^{2\tau+\beta-1}\left[\left(\frac{2\beta+\tau-\nu-1}{1+y}-\frac{2\alpha+\eta-\mu-1}{1-y}\right)(1-y^2)\frac{d}{dy}\right.$$
$$\left.+\frac{2\beta(\beta+\tau-1)}{1+y}+\frac{2\alpha(\alpha+\eta-1)}{1-y}-(\alpha+\beta)(\alpha+\beta+\eta+\tau-1)-n(n+\mu+\nu+1)\right]|P_n^{(\mu,\nu)}(y)\rangle \quad (A5)$$

Using the differential relation of the Jacobi polynomials (A2b) in the above, we obtain

$$\frac{1}{(\gamma\lambda)^2}\frac{d^2}{dx^2}|\phi_n(x)\rangle = A_n(1-y)^{2\eta+\alpha-1}(1+y)^{2\tau+\beta-1}\left\{\left[-(\alpha+\beta)(\alpha+\beta+\eta+\tau-1)-n(n+\mu+\nu+1)\right.\right.$$
$$\left.+\frac{2\beta(\beta+\tau-1)}{1+y}+\frac{2\alpha(\alpha+\eta-1)}{1-y}+\left(\frac{2\beta+\tau-\nu-1}{1+y}-\frac{2\alpha+\eta-\mu-1}{1-y}\right)\frac{2(\mu-\nu)n(n+\mu+\nu+1)}{(2n+\mu+\nu)(2n+\mu+\nu+2)}\right]|P_n^{(\mu,\nu)}\rangle$$
$$+2\left(\frac{2\beta+\tau-\nu-1}{1+y}-\frac{2\alpha+\eta-\mu-1}{1-y}\right)(n+\mu+\nu+1)$$
$$\left.\times\left[\frac{(n+\mu)(n+\nu)}{(2n+\mu+\nu)(2n+\mu+\nu+1)}|P_{n-1}^{(\mu,\nu)}\rangle-\frac{n(n+1)}{(2n+\mu+\nu+1)(2n+\mu+\nu+2)}|P_{n+1}^{(\mu,\nu)}\rangle\right]\right\} \quad (A6)$$

Now, we can evaluate the matrix elements of the kinetic energy operator $-\frac{1}{2}\frac{d^2}{dx^2}$ by projecting from left in (A6) with $\langle\phi_m|$ and integrating each term. For this integration, one should note that $\int_{x_-}^{x_+}dx=\int_{-1}^{+1}\frac{dy}{y'}$ where $y'$ is given by (A1) and the parameters are related as $2\alpha=\mu+\eta$ and $2\beta=\nu+\tau$. Consequently, we observe that all resulting integrals are of the form given by Eq. (11), which is evaluated in Appendix B. Nonetheless, one should take care in making the following change

$$A_n|P_{n-1}^{(\mu,\nu)}\rangle=\frac{A_n}{A_{n-1}}A_{n-1}|P_{n-1}^{(\mu,\nu)}\rangle=\sqrt{\frac{2n+\mu+\nu+1}{2n+\mu+\nu-1}\frac{n(n+\mu+\nu)}{(n+\mu)(n+\nu)}}A_{n-1}|P_{n-1}^{(\mu,\nu)}\rangle \quad (A7a)$$

$$A_n|P_{n+1}^{(\mu,\nu)}\rangle=\frac{A_n}{A_{n+1}}A_{n+1}|P_{n+1}^{(\mu,\nu)}\rangle=\sqrt{\frac{2n+\mu+\nu+1}{2n+\mu+\nu+3}\frac{(n+\mu+1)(n+\nu+1)}{(n+1)(n+\mu+\nu+1)}}A_{n+1}|P_{n+1}^{(\mu,\nu)}\rangle \quad (A7b)$$

Finally, we obtain the following matrix elements of the kinetic energy operator $-\frac{1}{2}\frac{d^2}{dx^2}$ in the Jacobi basis (9)

$$-\frac{4}{(\gamma\lambda)^2}\langle\phi_m|\left(-\frac{1}{2}\frac{d^2}{dx^2}\right)|\phi_n\rangle = -\left[\left(n+\frac{\mu+\nu+1}{2}\right)^2 + \frac{3}{4}(\eta+\tau-1)^2 + (\mu+\nu+1)(\eta+\tau-1)\right]F_{m,n}^{(\mu,\nu)}(2\eta-1,2\tau-1)$$

$$+\frac{1}{2}\left[(\mu+2\eta-1)^2 - (\eta-1)^2 - 4(2\eta-1)(\mu-\nu)G_n\right]F_{m,n}^{(\mu,\nu)}(2\eta-2,2\tau-1)$$

$$+\frac{1}{2}\left[(\nu+2\tau-1)^2 - (\tau-1)^2 + 4(2\tau-1)(\mu-\nu)G_n\right]F_{m,n}^{(\mu,\nu)}(2\eta-1,2\tau-2) \quad\text{(A8)}$$

$$+(n+\mu+\nu+1)D_{n-1}\left[(2\tau-1)F_{m,n-1}^{(\mu,\nu)}(2\eta-1,2\tau-2) - (2\eta-1)F_{m,n-1}^{(\mu,\nu)}(2\eta-2,2\tau-1)\right]$$

$$-nD_n\left[(2\tau-1)F_{m,n+1}^{(\mu,\nu)}(2\eta-1,2\tau-2) - (2\eta-1)F_{m,n+1}^{(\mu,\nu)}(2\eta-2,2\tau-1)\right]$$

$$+n\leftrightarrow m$$

where $G_n = \frac{n(n+\mu+\nu+1)}{(2n+\mu+\nu)(2n+\mu+\nu+2)}$ and $D_n$ is defined below Eq. (B2) in Appendix B. Note the addition of the $n\leftrightarrow m$ exchange in last step because the matrix $T$ is symmetric.

Now, if the coordinate transformation is such that one can write $(\lambda x)^2 = g(1-y)^p(1+y)^q$ then the kinetic energy operator can include the orbital term $\frac{\ell(\ell+1)}{2r^2}$ indicating that the problem is in 3D with spherical symmetry. In such case, we should add to the matrix (A8) the following

$$-\frac{4}{(\gamma\lambda)^2}\langle\phi_m|\frac{\ell(\ell+1)}{2x^2}|\phi_n\rangle = -\frac{2\ell(\ell+1)}{\gamma^2 g}\langle\phi_m|(1-y)^{-p}(1+y)^{-q}|\phi_n\rangle = -\frac{2\ell(\ell+1)}{\gamma^2 g}F_{m,n}^{(\mu,\nu)}(-p,-q) \quad\text{(A9)}$$

The result of adding the two matrices (A8) and (A9) is $-(2/\gamma\lambda)^2\langle\phi_m|T|\phi_n\rangle$. In Table 1, there is one case where this 3D problem appears. It corresponds to the fourth row where $(g,p,q) = (1,-1,1)$.

## Appendix B: Evaluating the integrals (11) and (15)

In this Appendix, we evaluate the integrals $F_{n,m}^{(\mu,\nu)}(\alpha,\beta)$ and $\tilde{F}_{n,m}^{(\mu,\nu)}(\alpha,\beta)$ defined in Eq. (11) and Eq. (15), respectively. We start with $F_{n,m}^{(\mu,\nu)}(\alpha,\beta)$ and make the following observations. The Jacobi polynomial identity $P_n^{(\mu,\nu)}(-y) = (-1)^n P_n^{(\nu,\mu)}(y)$ leads to the following exchange symmetry of the integral: $F_{n,m}^{(\mu,\nu)}(\alpha,\beta) = (-1)^{n+m}F_{n,m}^{(\nu,\mu)}(\beta,\alpha)$. The special case when both $\alpha$ and $\beta$ are non-negative integers could be easily evaluated by making repeated application of the three-term recursion relation (A2c). Thus, if $\alpha = k$ and $\beta = l$ then we obtain

$$F_{n,m}^{(\mu,\nu)}(k,l) = \left[(1-K)^k(1+K)^l\right]_{n,m}, \quad\text{(B1)}$$

where $K$ is the tridiagonal symmetric matrix whose elements are

$$K_{n,m} = C_n\delta_{n,m} + D_{n-1}\delta_{n,m+1} + D_n\delta_{n,m-1}, \quad\text{(B2)}$$

where $C_n = \frac{\nu^2-\mu^2}{(2n+\mu+\nu)(2n+\mu+\nu+2)}$ and $D_n = \frac{2}{2n+\mu+\nu+2}\sqrt{\frac{(n+1)(n+\mu+1)(n+\nu+1)(n+\mu+\nu+1)}{(2n+\mu+\nu+1)(2n+\mu+\nu+3)}}$. A special case is $F_{n,m}^{(\mu,\nu)}(0,0) = \delta_{n,m}$. For finite matrix calculation, it is worthwhile making the following comment about multiplication of finite versus infinite dimensional matrices. The multiplication of $N \times N$ truncated matrices $A$ and $B$ as $C^N = A^k B^l$ agrees with the infinite dimensional multiplication, $C^\infty$, only for the first $N - \lfloor \frac{k+l}{2} \rfloor$ rows and columns, where $\lfloor \frac{k+l}{2} \rfloor$ is the largest integer less than or equal to $\frac{k+l}{2}$. This should be taken into account when doing calculation using the matrix (B1), where $A = 1 - K$ and $B = 1 + K$.

Next, we base our general evaluation of $F_{n,m}^{(\mu,\nu)}(\alpha,\beta)$ on formula (2) of section 7.391 on top of page 807 in Ref. [11]. In the integral (11), we use the well-known expansion of $P_m^{(\mu,\nu)}(y)$ as

$$P_m^{(\mu,\nu)}(y) = \frac{(\mu+1)_m}{m!} \sum_{k=0}^{m} \frac{(-m)_k (m+\mu+\nu+1)_k}{(\mu+1)_k 2^k k!} (1-y)^k. \tag{B3}$$

Then the integral (11) becomes a sum of $m+1$ individual integrals each of which has the following form

$$\int_{-1}^{+1} (1-y)^{\mu+\alpha+k} (1+y)^{\nu+\beta} P_n^{(\mu,\nu)}(y) dy, \tag{B4}$$

where $k = 0, 1, 2, ..., m$. Now this integral looks like the one found in formula (2) cited above that could be written as

$$\int_{-1}^{+1} (1-y)^\rho (1+y)^\sigma P_n^{(\mu,\nu)}(y) dy = 2^{\rho+\sigma+1} \frac{\Gamma(\rho+1)\Gamma(\sigma+1)}{\Gamma(\rho+\sigma+2)} \frac{(\mu+1)_n}{n!} {}_3F_2\left(\begin{array}{c}-n, n+\mu+\nu+1, \rho+1\\ \mu+1, \rho+\sigma+2\end{array}\bigg|1\right), \tag{B5}$$

where $\rho > -1$ and $\sigma > -1$. We make the following map in this formula: $\sigma \mapsto \nu + \beta$, $\rho \mapsto \mu + \alpha + k$ and impose the constraints $\beta + \nu > -1$ and $\alpha + \mu > -k - 1$. If we name the integral (B4) as $I_k$ then using (B5) we can write

$$I_k = 2^{\mu+\nu+\alpha+\beta+k+1} \frac{\Gamma(\mu+\alpha+k+1)\Gamma(\nu+\beta+1)}{\Gamma(\mu+\nu+\alpha+\beta+k+2)} \frac{(\mu+1)_n}{n!} {}_3F_2\left(\begin{array}{c}-n, n+\mu+\nu+1, \mu+\alpha+k+1\\ \mu+1, \mu+\nu+\alpha+\beta+k+2\end{array}\bigg|1\right). \tag{B6}$$

Using the polynomial expansion (B3) and this result, we obtain

$$\begin{aligned}F_{n,m}^{(\mu,\nu)}(\alpha,\beta) &= \frac{A_n A_m}{\gamma} \frac{(\mu+1)_m}{m!} \sum_{k=0}^{m} \frac{(-m)_k (m+\mu+\nu+1)_k}{(\mu+1)_k 2^k k!} I_k \\ &= 2^{\mu+\nu+\alpha+\beta+1} \frac{\Gamma(\mu+\alpha+1)\Gamma(\nu+\beta+1)}{\Gamma(\mu+\nu+\alpha+\beta+2)} \frac{A_n A_m}{\gamma} \frac{(\mu+1)_n}{n!} \frac{(\mu+1)_m}{m!} \\ &\times \sum_{k=0}^{m} \frac{(\mu+\alpha+1)_k (-m)_k (m+\mu+\nu+1)_k}{(\mu+\nu+\alpha+\beta+2)_k (\mu+1)_k k!} {}_3F_2\left(\begin{array}{c}-n, n+\mu+\nu+1, \mu+\alpha+k+1\\ \mu+1, \mu+\nu+\alpha+\beta+k+2\end{array}\bigg|1\right) \end{aligned} \tag{B7}$$

Using the Pochhammer symbol identity $(a)_{n+m} = (a)_n (a+n)_m$ in the expansion of the ${}_3F_2$ series above, we can rewrite this as follows

$$F_{n,m}^{(\mu,\nu)}(\alpha,\beta) = 2^{\mu+\nu+\alpha+\beta+1} \frac{\Gamma(\mu+\alpha+1)\Gamma(\nu+\beta+1)}{\Gamma(\mu+\nu+\alpha+\beta+2)} \frac{A_n A_m}{\gamma} \frac{(\mu+1)_n}{n!} \frac{(\mu+1)_m}{m!}$$
$$\times \sum_{k=0}^{m}\sum_{l=0}^{n} \frac{(\mu+\alpha+1)_{k+l}(-m)_k(-n)_l(m+\mu+\nu+1)_k(n+\mu+\nu+1)_l}{(\mu+\nu+\alpha+\beta+2)_{k+l}(\mu+1)_k(\mu+1)_l\, k!\,l!} \quad (B8)$$

Now, since $F$ is a symmetric matrix (i.e., $F_{n,m}^{(\mu,\nu)} = F_{m,n}^{(\mu,\nu)}$), then for any given $n$, we only need to evaluate $F_{n,m}^{(\mu,\nu)}$ for $m=0,1,...,n$ (i.e., $m \le n$). As such, we would have calculated the lower triangle of the matrix while transposition ($F_{m,n}^{(\mu,\nu)} = F_{n,m}^{(\mu,\nu)}$ for $m > n$) gives the full matrix. We like to mention that the representation of $F_{n,m}^{(\mu,\nu)}(\alpha,\beta)$ given by Eq. (B8) is numerically unsuitable for large values of $n$ and $m$. It is our conjecture that a suitable representation would by a single finite sum (from zero to $n \wedge m$) of products of two $_3F_2$ functions.

Now, we evaluate the integral $\tilde{F}_{n,m}^{(\mu,\nu)}(\alpha,\beta)$ defined by Eq. (15). Using (B3), we can write

$$\frac{d}{dy}P_m^{(\mu,\nu)}(y) = -\frac{(\mu+1)_m}{m!}\sum_{k=0}^{m-1}\frac{(-m)_{k+1}(m+\mu+\nu+1)_{k+1}}{(\mu+1)_{k+1}\, 2^{k+1} k!}(1-y)^k, \quad (B9)$$

which is zero for $m=0$. Proceeding as above and with the help of the integral (B5), we obtain

$$\tilde{F}_{n,m}^{(\mu,\nu)}(\alpha,\beta) = -2^{\mu+\nu+\alpha+\beta} \frac{\Gamma(\mu+\alpha+1)\Gamma(\nu+\beta+1)}{\Gamma(\mu+\nu+\alpha+\beta+2)} \frac{A_n A_m}{\gamma} \frac{(\mu+1)_n}{n!} \frac{(\mu+1)_m}{m!}$$
$$\times \sum_{k=0}^{m-1}\sum_{l=0}^{n} \frac{(\mu+\alpha+1)_{k+l}(-m)_{k+1}(-n)_l(m+\mu+\nu+1)_{k+1}(n+\mu+\nu+1)_l}{(\mu+\nu+\alpha+\beta+2)_{k+l}(\mu+1)_{k+1}(\mu+1)_l\, k!\,l!} \quad (B10)$$

Note that $\tilde{F}_{n,m}^{(\mu,\nu)}(\alpha,\beta) \ne \tilde{F}_{m,n}^{(\mu,\nu)}(\alpha,\beta)$ and $\tilde{F}_{n,0}^{(\mu,\nu)}(\alpha,\beta) = 0$.

**Remark**: An alternative evaluation of the integral (11) is by using the linearization formula of the product of two Jacobi polynomials (see, for example equations (1.2), (1.3), (1.5) and (1.6) in Ref. [12])

$$P_n^{(\mu,\nu)}(y)P_m^{(\mu,\nu)}(y) = \sum_{k=|n-m|}^{n+m} f_{n,m}^k P_k^{(\mu,\nu)}(y), \quad (B11)$$

where $f_{n,m}^k$ depends on $\mu$ and $\nu$. Then, using formula (B5) above results in the following value for the integral

$$2^{\mu+\nu+\alpha+\beta+1}\frac{\Gamma(\mu+\alpha+1)\Gamma(\nu+\beta+1)}{\Gamma(\mu+\nu+\alpha+\beta+2)}\sum_{k=|n-m|}^{n+m} f_{n,m}^k \frac{(\mu+1)_k}{k!}\,_3F_2\!\left(\begin{array}{c}-k,k+\mu+\nu+1,\mu+\alpha+1\\ \mu+1,\mu+\nu+\alpha+\beta+2\end{array}\bigg|1\right). \quad (B12)$$

Unfortunately, the coefficient $f_{n,m}^k$ is too complicated relative to (B8) for an efficient calculation since it involves sums of higher order hypergeometric functions (see, Eq. (1.5) and Eq. (1.6) in [12] and Eq. (10) in [13]). Additionally, a recursion relation satisfied by $\{f_{n,m}^k\}$ is not known [14]. Has such recursion been available, we could have obtained all of them simply by calculating only the first few.

## Table Caption

**Table 1**: The parameters of orthonormal Jacobi bases given by Eq. (9) that correspond to the examples in section 2 with $2\alpha = \mu + \eta$ and $2\beta = \nu + \tau$. The normalization constant is $A_n = \sqrt{\gamma}\sqrt{\frac{2n+\mu+\nu+1}{2^{\mu+\nu+1}} \frac{\Gamma(n+1)\Gamma(n+\mu+\nu+1)}{\Gamma(n+\mu+1)\Gamma(n+\nu+1)}}$. All coordinate transformations $y(x)$ satisfy $dy/dx = \lambda\gamma(1-y)^\eta (1+y)^\tau$.

## Figures Captions

**Fig. 1**: (a) The potential component $\tilde{V}(x)$ for the 1D system described in subsection 2.1 shown as a solid curve for $b+a=7$ and $c+a=2$. A perfect match with the function $\tilde{V}(x) = \tilde{V}_0 + \frac{\tilde{V}_1}{\cosh^2(\lambda x)}$ is shown by the dotted curve for a given fitting parameters $\tilde{V}_0$ and $\tilde{V}_1$. (b) Plot of the total potential function (18) for a given equivalence parameters $q_0$ and $q_1$ where we also superimpose the bound states energy levels.

**Fig. 2**: (a) The potential component $\tilde{V}(x)$ for the 1D system described in subsection 2.2 shown as a solid curve for $b+a=6$ and $c+a=1$. A perfect match with the function $\tilde{V}(x) = \tilde{V}_0 + \tilde{V}_1 \tanh(\lambda x) + \frac{\tilde{V}_2}{\cosh^2(\lambda x)}$ is shown by the dotted curve for a given set of fitting parameters $\{\tilde{V}_i\}$. (b) Plot of the total potential function (20) for a given equivalence parameters $q_0$ and $q_1$ where we also superimpose the bound states energy levels.

**Fig. 3**: (a) The potential component $\tilde{V}(x)$ for the 1D system described in subsection 2.3 for $b+a=20$ and $c+a=2$. (b) Plot of the total potential function (22) for a given equivalence parameters $q_0$ and $q_1$ where we also superimpose the bound states energy levels.

**Fig. 4**: The radial potential function $V(r)$ plus the orbital term $\frac{\ell(\ell+1)}{2r^2}$ for the 3D system with spherical symmetry described in subsection 2.4 shown as a solid curve for $\ell=2$, $c+a=3$ and $b+a = \ell + \frac{3}{2}$. A perfect match with the function $V(r) = V_0 + \frac{V_1}{1+(\lambda r)^2} + \frac{V_2}{\left[1+(\lambda r)^2\right]^2}$ is shown by the dotted curve for a given set of fitting parameters $\{V_i\}$ and equivalence parameters $\{q_0, q_1\}$ where we also superimpose the bound states energy levels.

**Fig. 5**: (a) The potential component $\tilde{V}(x)$ for the 1D system described in subsection 2.5 shown as a solid curve for $b+a=5$ and $c+a=1$. A perfect match with the function $\tilde{V}(x) = \tilde{V}_0 [1-\sin(\lambda x)]$ is shown by the dotted curve for a given fitting parameter $\tilde{V}_0$. (b) Plot of the total potential function (28) with the equivalence parameters $q_0 = 1$ and $q_1 = 0$.

**Table 1**

| $y(x)$ | $x$ | $\eta$ | $\tau$ | $\gamma$ |
|---|---|---|---|---|
| $2\tanh^2(\lambda x)-1$ | $x \geq 0$ | $1$ | $\frac{1}{2}$ | $\sqrt{2}$ |
| $\tanh(\lambda x)$ | $-\infty < x < +\infty$ | $1$ | $1$ | $1$ |
| $1-2e^{-\lambda x}$ | $x \geq 0$ | $1$ | $0$ | $1$ |
| $\dfrac{(\lambda x)^2-1}{(\lambda x)^2+1}$ | $x \geq 0$ | $\frac{3}{2}$ | $\frac{1}{2}$ | $1$ |
| $\sin(\lambda x)$ | $-\frac{\pi}{2} \geq \lambda x \geq +\frac{\pi}{2}$ | $\frac{1}{2}$ | $\frac{1}{2}$ | $1$ |

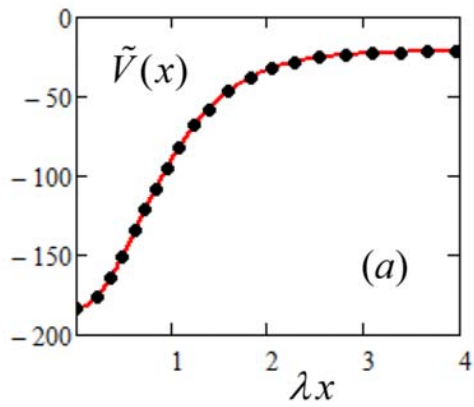
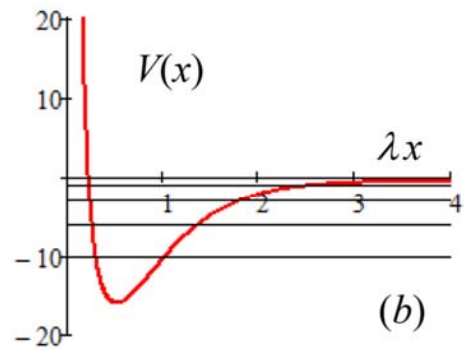

**Fig. 1**

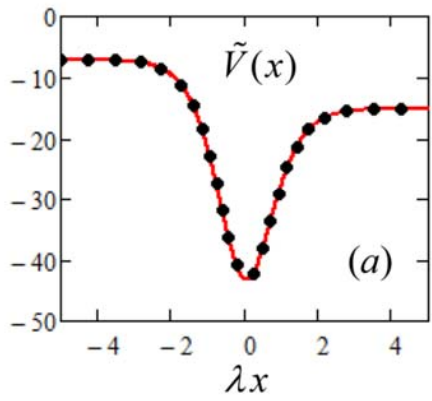 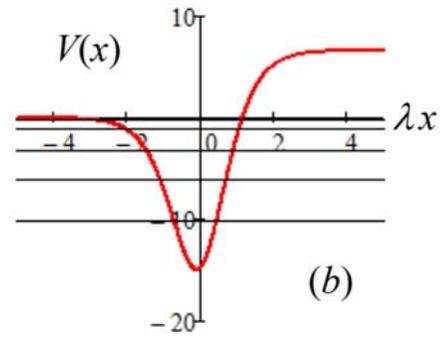

**Fig. 2**

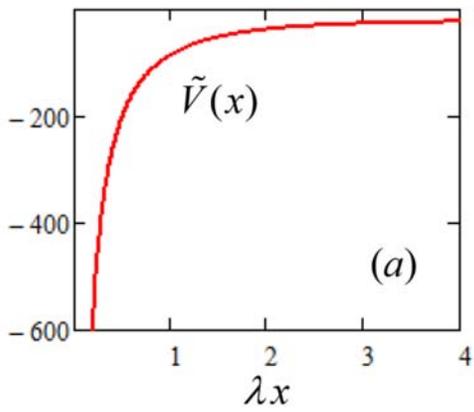 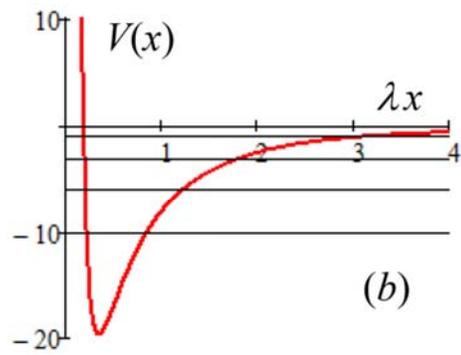

**Fig. 3**

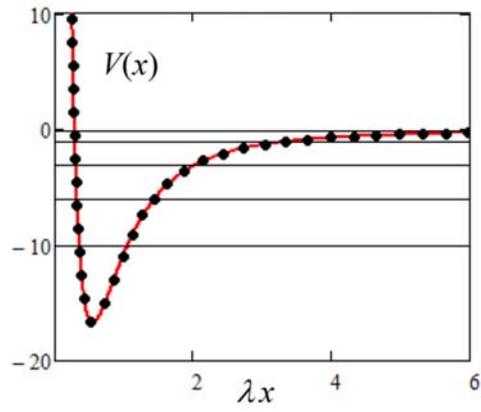

**Fig. 4**

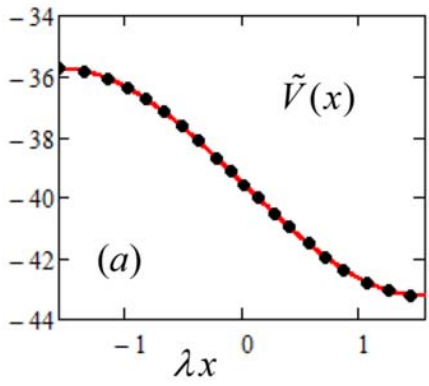
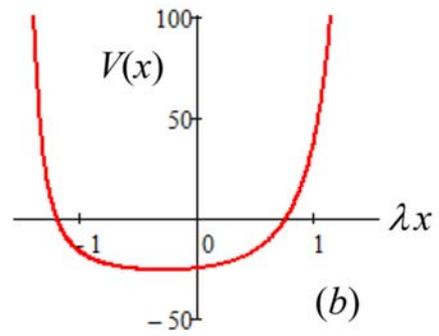

**Fig. 5**